\documentclass[nofootinbib, amsmath,amssymb, aps,]{revtex4}

\usepackage{hyperref}

\usepackage{graphicx}
\usepackage{dcolumn}
\usepackage{bm}

\usepackage{booktabs}
\usepackage{topcapt}

\usepackage{xcolor}

\usepackage{slashed} 

 \usepackage{ulem}
 \usepackage{cancel}
 
 \usepackage{calligra}

\begin{document}
\title{Infrared divergences and the photon mass in QED}
\author{Orlando Oliveira}

\affiliation{CFisUC, Departament of Physics, University of Coimbra, 3004-516 Coimbra, Portugal}

\begin{abstract}
The infrared properties of QED are investigated within the framework of the Dyson-Schwinger equations. 
Our study finds that, independently of the value of the coupling constant,
requiring the photon self-energy to be finite for any momenta, 
combined with a smooth behavior for the photon-fermion vertex, 
is equivalent to state that the photon is massless and that the photon propagator diverges at low momenta as $1/k^2$.
Furthermore, the Schwinger mechanism to generate, in a gauge invariant way, a photon mass is investigated and the form factors that can 
be at the origin of a possible photon mass are identified. 
For the Schwinger mechanism the link between the finiteness of the photon self-energy and the masslessness of the photon is lost.
The infrared behavior of the fermion gap equation and the vertex equation are found to be infrared safe 
integral equations. Moreover, by studying chiral fermions within QED it is observed that the requirement of the finiteness of the photon self-energy 
translates into a fermion propagator that behaves as $\slashed{p}/p^4$.
\end{abstract}

\maketitle
\tableofcontents

\section{Introduction and Motivation}

Quantum Electrodynamics (QED) has guided our understanding of quantum field theory (QFT). 
The construction of the perturbative solutions in QED encounters infinities arising from both the ultraviolet (UV) and the infrared (IR) regimes. 
The treatment of ultraviolet divergences led to the development of renormalization, that is a cornerstone of modern QFT. 
However, the nature of IR divergences is quite different from that of their UV counterparts. 
These IR infinities are linked to the photon being a massless particle. 
The current experimental upper limit on the photon mass, as listed in the particle data book \cite{PDGphotonmass,ParticleDataGroup:2024cfk}, is $m_\gamma < 10^{-18}$ eV,
 making it reasonable to assume that $m_\gamma = 0$. Furthermore,
 QED is an Abelian gauge theory, and gauge symmetry prevents the inclusion of an explicit mass term for the photon in the 
 Lagrangean. The reasoning on an explicit mass term in the Lagrangean applies also to the gauge sector of non-Abelian theories. As long as U(1) gauge symmetry is not 
 a broken symmetry,  the photon remains massless.

In the perturbative Feynman diagrammatic expansion for QFT, the presence of zero-mass particles leads to low-momentum divergences.
These can be prevented by replacing massless propagators with massive ones; that is, by assigning a fictitious small mass to the massless particles.
Typically, the infrared problem associated with the presence of massless particles manifests as an infinite contribution to certain cross-sections.
In QED, the computation of the Bremsstrahlung cross-section to lowest order in the coupling constant is a textbook example of this phenomenon.

If the photon is indeed massless, then adding any number of zero-frequency photons $| \gamma \rangle_0$ to a Fock-space state $| \psi \rangle$,
results in a new state $| \psi^\prime \rangle = | \psi \rangle \otimes |n \,  \gamma \rangle_0$ that is degenerate in energy.
Consequently, conventional Fock-space states are ill-defined. 
Furthermore, these infrared divergences imply that all $S$-matrix elements between conventional Fock-space states vanish.

The infrared divergences in the computation of cross-sections between Fock-space states can be resolved by considering inclusive cross-sections.
Summing over all possible final states hides the IR divergences beneath the experimental resolution 
\cite{Bloch:1937pw,Yennie:1961ad,Kinoshita:1962ur,Lee:1964is}, yielding a finite transition probability.
Alternatively, one can replace the conventional Fock-space basis states with dressed states, where the Fock-space states are surrounded by clouds of soft photons. This approach solves the problem of IR divergences in QED \cite{Chung:1965zza,Kibble:1968oug,Kibble:1968npb,Kibble:1968lka,Kulish:1970ut} but increases the complexity of any calculation.
The introduction of infinitely degenerate vacuum states can also properly handle the infrared divergences of QED, provided one accounts for transitions between these different vacuum states \cite{Kapec:2017tkm,Strominger:2017zoo}.
The  infrared and ultraviolet divergences can be handled simultaneously within 
Bogolubov–Parasuk–Hepp–Zimmermann (BPHZ) renormalization procedure that relies on the $R^\star$-operation,
see \cite{Chetyrkin:1982nn,Chetyrkin:1984xa,Herzog:2017ohr} and references therein, and render finite probability transitions.

Our objective is to discuss the infrared divergences in QED from the perspective of the Dyson-Schwinger equations (DSEs).
The DSEs form an infinite tower of integral equations that relate all QED Green functions. For practical reasons, one must consider a truncated version of the full set.
In this work, we consider only the equations for the fermion propagator (the gap equation), the photon propagator (the photon gap equation),
and the photon-fermion one-particle irreducible Green function (photon-fermion vertex) integral equation.
The discussion applies to a general linear covariant gauge, identified by the gauge-fixing parameter $\xi$.
The integral equations will not be derived here, but the interested reader can find a derivation in \cite{Oliveira:2022bar}, whose notation we adopt, or in, for example, \cite{Roberts:1994dr}.
As our starting point, we assume that the DSEs and the self-energies are finite. This is always the case before taking the limit $i \epsilon \rightarrow 0$, and
all manipulations should be understood as being performed with a small but non-vanishing $\epsilon$.
We emphasize that our analysis does not require solving any of the integral equations.

As will be discussed, the photon gap equation does not lead necessarily to an infrared finite photon propagator. However, the requirement of finiteness for the photon 
self-energy, together with a smooth behavior for the photon-fermion vertex, is equivalent to state that the photon is necessarily a massless particle. Furthermore, the photon 
propagator DSE implies also that the photon propagator diverges as $1/k^2$ for small photon momentum $k$. Thus, the requirement of finiteness for the self-energy
links the infrared properties of QED, the masslessness of the photon, and the divergence of the photon propagator at low momentum.
In addition to conventional QED, we investigate the Schwinger mechanism \cite{Schwinger:1962tn,Schwinger:1962tp} for photon mass generation, which can occur via a singular 
behavior in the transverse component of the photon-fermion vertex. We show that certain transverse components of the vertex can indeed generate  a finite and non-vanishing photon 
mass, and we identify the specific components responsible for this effect via the Schwinger mechanism.
However, within the Schwinger mechanism the photon is not necessarily a massless particle.
Regarding the photon mass, we remind the reader that, as discussed in \cite{Oliveira:2025ftr}, conventional perturbation theory results in a photon mass 
that is proportional to the fermion mass that appears in the loops.

The fermion gap equation is also investigated and our conclusion is that it is infrared safe. The particular case of QED with chiral fermions is also studied. 
From the finiteness of the photon self-energy, the corresponding fermion propagator properties are derived. 
Surprisingly, it follows that the chiral fermion propagator exhibits an improved UV behavior compared to the conventional perturbative solution.

The infrared properties of the DSE for the photon-fermion vertex are investigated by applying the same type of reasoning, enabling us to prove that the vertex is infrared finite,
and is compatible with our initial assumption.
All the results described, with the exception of those associated with the Schwinger mechanism, assume a smooth vertex behavior and require the DSEs to be finite. 

The analysis performed in the current work and its conclusions are independent of the value of the QED coupling constant. 
The results derived suggest that the typical infrared problems of perturbation theory are a limitation of the perturbative framework itself, rather than an intrinsic property of QED.

The generalization of the results derived in this work to non-Abelian gauge theories, and in particular to QCD, requires the consideration of additional equations, see e.g.
\cite{Pascual:1984zb,Alkofer:2000wg,Fischer:2006ub,Huber:2018ned} and references therein.
In QCD, besides fermions and gluons, one must also account for non-vanishing three- and four-gluon one-particle irreducible Green functions, as well as ghost-related 
Green functions. Furthermore, in QCD the photon-fermion Ward-Takahashi identity (WTI) is replaced by a Slavnov-Taylor identity, whose formal solution 
\cite{Aguilar:2010cn,Oliveira:2018fkj} requires contributions from the quark-ghost scattering kernel. 
An equivalent analysis is necessarily more complex and is left for future work.
The non-Abelian version of the Schwinger mechanism seems to play an important role in gluon dynamics
\cite{Aguilar:2016ock,Aguilar:2023mdv,Ferreira:2025anh}
and has been invoked to recover, within the continuum formalism of QCD, the lattice results \cite{Bogolubsky:2009dc,Duarte:2016iko} for the two-point functions.

This work is organized as follows. In Sec. \ref{Sec:Intro} we introduce the definitions used for the two point functions, for the photon-fermion vertex and provide
the renormalized DSE. In \ref{Sec:vertice-tensor} the photon-fermion vertex is described in detail, and a  tensor basis for this Green function is introduced.
Further, the longitudinal form factors are given in terms of the fermion propagator functions as solutions of the vertex Ward-Takahashi identity.
In Sec. \ref{Sec:PhotonGapEq}, the photon gap equation is investigated together with the conditions for a finite DSE and self-energy, that include the photon mass and the
behavior of the photon propagator at low momenta. In \ref{Sec:fermions} the fermion gap equation and its infrared properties are investigated.
The case of chiral fermions is also studied. The vertex and its infrared properties are studied in
Sec. \ref{Sec:VertEq}. The gauge invariant Schwinger mechanism of mass generation for the photon is discussed in \ref{Sec:Schwinger}. In particular,
the contribution of the various transverse form factors to a possible photon mass is worked out. Finally, in \ref{Sec:Conclusions} we summarize and conclude.

\section{Definitions and Integral Equations \label{Sec:Intro}}

In this section the notation will be defined and the renormalized DSEs introduced. In all cases, the equations are written in Minkowski spacetime.

\subsection{The propagators and the photon-fermion vertex}

For  linear covariant gauges, the photon propagator in momentum space reads
\begin{equation}
    D_{\mu\nu}(k) = - \left( g_{\mu\nu} - \frac{k_\mu k_\nu}{k^2} \right) \, D(k^2) - \frac{\xi}{k^2} \, \frac{k_\mu k_\nu}{k^2} 
    = - P^\perp_{\mu\nu} (k) \, D(k^2) - \frac{\xi}{k^2} \, P^L_{\mu \nu}(k)  \ .
    \label{Def:PhotonProp}
\end{equation}
The Lorentz invariant scalar function $D(k^2)$, also referred as the photon propagator, is in tree-level perturbation theory given by
$D(k^2) = 1 / ( k^2 + i \, \epsilon)$. The inverse of the fermion propagator is 
\begin{equation}
 S^{-1}(p) = A(p^2) \, \slashed{p} - B(p^2) + i \, \epsilon \ ,
\end{equation}
where $A(p^2)$ and $B(p^2)$ are Lorentz scalar functions and the limit $\epsilon \rightarrow 0^+$ is to be taken at the end of the calculations.
In the following, unless clearly stated, the $i \epsilon$ term will be omitted from now on.

The photon-fermion one-particle irreducible Green function (1PI), also named photon-fermion vertex,
 will be written as $\Gamma^\mu (p, -p-k; k)$, where $p$ is the incoming fermion
momentum, $p+k$ is the outgoing fermion momentum and $k$ the incoming photon momentum. This Green function has to comply with
the symmetries of QED and, in particular, with gauge symmetry that relates its longitudinal part, relative to the photon momentum, to the fermion propagator 
via a Ward-Takahashi identity (WTI). The Ward-Takahashi vertex identity determines \cite{Ball:1980ay} (not-uniquely) the longitudinal part of $\Gamma^\mu$. 
We postpone the discussion on the vertex to a later section.

\subsection{The QED Dyson-Schwinger Equations}

The fields to be considered in QED are the fermion $\psi$ and the photon $A_\mu$ fields. Bare  and physical quantities are related by renormalization constants
that can be defined as
\begin{eqnarray}
& &
   A_\mu  = Z^{\frac{1}{2}}_3 \, A^{(phys)}_\mu   , \quad
   \psi  = Z^{\frac{1}{2}}_2 \, \psi^{(phys)} , \quad
   g = \frac{Z_1}{Z_2 \, Z^{\frac{1}{2}}_3}      g^{(phys)}    ,  
\quad
   m = \frac{Z_0}{Z_2} \, m^{(phys)} \quad\mbox{and}\quad
  \xi = Z_3 \,  \xi^{(phys)}  ,
\end{eqnarray}
where $g$ is the coupling constant and $\xi$ is the gauge fixing parameter that defines the linear covariant gauges. Further details on QED can be seen in \cite{Oliveira:2022bar},
whose notation is followed.
The Ward-Takahashi identity for the fermion-photon vertex requires $Z_1 = Z_2$, see e.g. \cite{Roberts:1994dr,Oliveira:2022bar} and references therein, reducing the
number of independent $Z_i$ to be determined.

The renormalized integral equations to be considered in this work are the fermion gap equation
\begin{eqnarray}
  S^{-1} (p)  & = &  Z_2 \,   \slashed{p} - Z_0 \, m 
     - \, i \, g^2  \, Z_2 \, \int \frac{d^4 k}{( 2 \, \pi )^4} ~   D_{\mu\nu} (k) ~  \bigg[ \gamma^\mu ~ S(p - k) ~   {\Gamma}^\nu (p-k, - p; k)  \bigg]   \ ,
     \label{DSE-fermion-gap}
\end{eqnarray}
the  photon gap equation 
\begin{eqnarray}
  \frac{1}{D(k^2)}  & = &
  Z_3 \,  k^2 
  - i \, \frac{g^2}{3} \,  Z_2 \,  \int \frac{d^4 p}{(2 \, \pi)^4} ~ \text{Tr} \bigg[ \gamma_\mu \, S(p) \, {\Gamma}^\mu(p, -p + k; -k ) \, S(p - k ) \bigg] \ ,
     \label{DSE-photon-gap}
\end{eqnarray}
together with the equation for the one-particle irreducible photon-fermion Green function that reads
\begin{eqnarray}
& & 
   {\Gamma}^\mu (p, \, -p -k; \, k)  ~  =  ~  Z_2 \, \gamma^\mu  ~ + ~ i \, g^2 \,  Z_2 \, \int \frac{d^4q}{(2 \, \pi)^4} ~D_{\zeta\zeta^\prime}(q) \nonumber \\
   & & \hspace{1cm}
    \Bigg\{
                                \gamma^\zeta ~ S(p-q) ~ {\Gamma}^\mu ( p - q, \, -p -k + q; \, k) ~
                                S(p+k-q) ~ {\Gamma}^{\zeta^\prime} (p +k-q, \, -p-k; q) 
     \nonumber \\
    & & \hspace{8cm} 
                 ~ + ~
                                \gamma^\zeta \, S(p-q) \, {\Gamma}^{\zeta^\prime\mu} (p - q, \, -p-k; \, q , \, k) 
                                \Bigg\}  \, .
                                \label{DSE-vertex}
\end{eqnarray}
${\Gamma}^{\mu\nu}$ is the one-particle irreducible two-photon-two-fermion Green function and is the solution of a DSE  that
call for Green functions with larger number of external legs. No further DSE will be considered in the current work, besides those already stated.
The three independent renormalization constants are fixed by the conditions
\begin{equation}
A(\mu^2_F) = 1, \qquad B(\mu^2_F) = m \qquad\mbox{ and }\qquad D(\mu^2_B) = \frac{1}{\mu^2_B}  \ ,
  \label{Renormalization-Conditions}
\end{equation}
where $\mu_F$ and $\mu_B$ are the renormalization mass scales for the fermion and the boson fields.

\section{The  photon-fermion vertex \label{Sec:vertice-tensor}}

For studying the photon-fermion vertex $\Gamma^\mu$ a tensor basis of operators will be introduced. A basis for this Green function
that is compatible with the symmetries of QED requires twelve scalar form factors \cite{Ball:1980ay}. The photon-fermion vertex can be 
written as
\begin{equation}
 \Gamma^\mu (p_2, \, p_1; \, p_3)  = \Gamma^\mu_L (p_2, \, p_1; \, p_3) +  \Gamma^\mu_T (p_2, \, p_1; \, p_3) \ ,
\end{equation}
where $p_2$ is the incoming fermion momentum, $-p_1$ is the outgoing fermion momentum, $p_3$ is the incoming photon momentum,
$\Gamma^\mu_T$ is the transverse component, relative to the photon momentum, of the vertex and complies with 
$p^\mu_3 ~ \Gamma_{T \, \mu}(p_2, \, p_1; \, p_3) = 0$.
$\Gamma^\mu_L$ is the longitudinal component of the vertex. From the momenta convention it follows that  $p_1 + p_2 + p_3 = 0$. 
The longitudinal and transverse components of the photon-fermion vertex are, respectively, 
 \begin{eqnarray}
 \Gamma_{L \, \mu} (p_2, \, p_1; p_3) & = & \sum^4_{i=1} \lambda_i (p^2_1, \, p^2_2, \, p^2_3) \, L^{(i)}_\mu (p_1, \, p_2, \,  p_3) \ ,  
 \label{Eq:photon-fermion_vertex-longitudinal}  \\
  \Gamma_{T \, \mu} (p_2, \, p_1; p_3) & = &  \sum^8_{i=1} \tau_i (p^2_1, \, p^2_2, \, p^2_3) \, T^{(i)}_\mu (p_1, \, p_2, \,  p_3) \ ,
 \label{Eq:photon-fermion_vertex-transverse}
\end{eqnarray}
where $L^{(i)}_\mu$ and $T^{(i)}_\mu$ are the basis of tensor operators to be considered, and $\lambda_i$ and $\tau_i$
are Lorentz scalar functions of the momenta. Note the different ordering of the momenta in the l.h.s. and r.h.s in Eqs. (\ref{Eq:photon-fermion_vertex-longitudinal})
and  (\ref{Eq:photon-fermion_vertex-transverse}). For the longitudinal part of the vertex the Ball-Chiu longitudinal set of operators \cite{Ball:1980ay} 
\begin{eqnarray}
L^{(1)}_\mu (p_1, \, p_2, \,  p_3) & = & \gamma_\mu \ ,  \label{TensorBasis-L1} \\
L^{(2)}_\mu (p_1, \, p_2, \,  p_3) & = & \big( \slashed{p}_1 - \slashed{p}_2 \big) \big( p_{1} - p_{2} \big)_\mu \ , \\
L^{(3)}_\mu (p_1, \, p_2, \,  p_3) & = & \big( p_{1} - p_{2} \big)_\mu \ , \\
L^{(4)}_\mu (p_1, \, p_2, \,  p_3) & = & \sigma_{\mu\nu} \big( p_{1} - p_{2} \big)^\nu \ ,
\label{TensorBasis-Long} 
\end{eqnarray}
will be considered, while for the transverse component the  K{\i}z{\i}lersu-Reenders-Pennington basis of operators \cite{Kizilersu:1995iz} 
\begin{eqnarray}
T^{(1)}_\mu (p_1, \, p_2, \,  p_3) & = & p_{1 \, _\mu} \big( p_2 \cdot p_3 \big) - p_{2 \, _\mu}  \big( p_1 \cdot p_3 \big)  \ , \label{TensorBasis-T1} \\
T^{(2)}_\mu (p_1, \, p_2, \,  p_3) & = & - \, T^{(1)}_\mu (p_1, \, p_2, \,  p_3) ~ \big( \slashed{p}_1 - \slashed{p}_2 \big)  \ ,  \label{TensorBasis-T2} \\
T^{(3)}_\mu (p_1, \, p_2, \,  p_3) & = & p^2_3 \, \gamma_\mu - p_{3 \, _\mu} \, \slashed{p}_3  \ ,  \label{TensorBasis-T3} \\
T^{(4)}_\mu (p_1, \, p_2, \,  p_3) & = & T^{(1)}_\mu (p_1, \, p_2, \,  p_3) ~ \sigma_{\alpha\beta} \, p^\alpha_1 \,  p^\beta_2 \ ,  \label{TensorBasis-T4} \\
T^{(5)}_\mu (p_1, \, p_2, \,  p_3) & = & \sigma_{\mu\nu} \, p^\nu_3 \ ,  \label{TensorBasis-T5} \\
T^{(6)}_\mu (p_1, \, p_2, \,  p_3) & = & \gamma_\mu \big( p^2_1 - p^2_2 \big) + \big( p_{1} - p_{2} \big)_\mu \, \slashed{p}_3 \ ,  \label{TensorBasis-T6} \\
T^{(7)}_\mu (p_1, \, p_2, \,  p_3) & = &  - \, \frac{1}{2} \, \big( p^2_1 - p^2_2 \big) \, \big[ \gamma_\mu \,  \big( \slashed{p}_1 - \slashed{p}_2 \big)  - \big( p_{1} - p_{2} \big)_\mu\big] 
               - \big( p_{1} - p_{2} \big)_\mu ~ \sigma_{\alpha\beta} \, p^\alpha_1 \,  p^\beta_2 \ , \label{TensorBasis-T7}  \\
T^{(8)}_\mu (p_1, \, p_2, \,  p_3) & = &  - \, \gamma_\mu \, \sigma_{\alpha\beta} \, p^\alpha_1 \,  p^\beta_2  \, + \, p_{1 \, _\mu} \slashed{p}_2 \, - \,  p_{2 \, _\mu} \slashed{p}_1  \ ,
 \label{TensorBasis-T8} \label{TensorBasis-Ortho} 
\end{eqnarray}
that is free of kinematical singularities, will be used. 
In Eqs (\ref{TensorBasis-L1}) to (\ref{TensorBasis-Ortho}) we take the definition 
$\sigma_{\mu\nu} = \frac{1}{2} \, [ \gamma_\mu \, , \, \gamma_\nu ]$. 

The longitudinal form factors $\lambda_i$'s can be written in terms of the fermion propagator functions $A$ and $B$ with the help of the vertex WTI, 
see e.g. \cite{Ball:1980ay} and \cite{Oliveira:2022bar}, and they read
\begin{eqnarray}
  \lambda_1 (p^2_1, \, p^2_2, \, p^2_3) & = & \frac{1}{2} \bigg( A\big( p^2_1 \big)  + A\big(p^2_2\big) \bigg)  \ ,    \label{EQ:L1} \\
  \lambda_2 (p^2_1, \, p^2_2, \, p^2_3) & = & \frac{1}{2 \, \big( p^2_1 - p^2_2 \big)}   \bigg( A\big( p^2_1 \big)  -  A\big(p^2_2\big) \bigg) \ , \label{EQ:L2} \\
  \lambda_3 (p^2_1, \, p^2_2, \, p^2_3) & = & \frac{1}{ p^2_1 - p^2_2 }   \bigg( B\big(p^2_1\big) - B\big( p^2_2 \big)    \bigg) \ ,   \label{EQ:L3} \\
  \lambda_4 (p^2_1, \, p^2_2, \, p^2_3) & = & 0 \ .  \label{EQ:L4}
\end{eqnarray}
Assuming that $A$ and $B$ are smooth functions, then $\lambda_2$ and $\lambda_3$ are regular in the limit of $p^2_1 \rightarrow p^2_2$ and are 
proportional to the derivatives of $A$ and $B$, respectively, at $p^2_1 = p^2_2 = p^2$.  For zero photon momentum, assuming a smooth behavior of all form factors,
it comes that
\begin{equation}
  \lambda_1 (p^2, \, p^2, 0) = A(p^2) \ , \qquad
  \lambda_2 (p^2, \, p^2, 0) = \frac{1}{2} \,  \frac{d A(p^2)}{d p^2} \qquad\mbox{ and }\qquad
  \lambda_3 (p^2, \, p^2, 0) = \frac{d B(p^2)}{d p^2} \ 
  \label{LongVertex-ZeroMom}
\end{equation}
in agreement with the Ward identity \cite{Ward:1950xp} for the vertex.

For zero photon momentum the operators in the K{\i}z{\i}lersu-Reenders-Pennington basis all vanish and, unless the form factors $\tau_i$ are singular at
this kinematical point, the one-particle irreducible Green function $\Gamma^\mu(p, \, -p; \, 0)$ is completely described by $\Gamma_L$, whose  form factors 
$\lambda_i$ are given in Eqs (\ref{LongVertex-ZeroMom}) in terms of the fermion propagator functions $A$ and $B$.

\section{The photon gap equation \label{Sec:PhotonGapEq}}

By describing the photon-fermion vertex $\Gamma^\mu$ with the above tensor basis,  handling the Dirac algebra with the help of FeynCalc
\cite{FeynCalc1,FeynCalc2,FeynCalc3},  the integral equation for the photon propagator (\ref{DSE-photon-gap}) become
\begin{eqnarray}
& &
  \frac{1}{D(k^2)}  = 
   k^2 \Bigg\{ Z_3 \, 
  - i \, \frac{g^2}{3} \, Z_2 \,  \int \frac{d^4 p}{(2 \, \pi)^4} ~  \frac{1}{A^2\left(\left(p + \frac{k}{2}\right)^2\right) \, (p + \frac{k}{2})^2 - B^2\left(\left(p + \frac{k}{2}\right)^2\right)} 
  \nonumber \\
  & & \hspace{5.2cm}
                                                                             ~ \frac{1}{A^2\left(\left(p - \frac{k}{2}\right)^2\right) \, (p-\frac{k}{2})^2 - B^2\left(\left(p - \frac{k}{2}\right)^2\right)}  \nonumber \\
  & & 
   \Bigg\{ ~
A\left(\left(p + \frac{k}{2}\right)^2\right) A\left(\left(p - \frac{k}{2}\right)^2\right) 
\Bigg[
 2 \, \lambda_1 \,  \bigg( 1 - 4 \, \frac{p^2}{k^2} \bigg)
 ~ + ~  4 \, \lambda_2   \bigg(  \frac{ 4 \, p^4 - 2 \, (pk)^2}{ k^2} + p^2 \bigg)
 \nonumber \\
 & & \hspace{6cm}  
 + ~  2 \, \tau_2  \,  \bigg(  - 4 \, \frac{p^2  (pk)^2}{k^2} + k^2 p^2 + 4 \, p^4 - (pk)^2 \bigg)
   \nonumber \\
   & & \hspace{6cm}                    
   + ~  \, \tau_3 \,  \bigg( - 8 \, \frac{(pk)^2}{k^2} + 3 \, k^2 - 4 \, p^2 \bigg)
   \nonumber \\
   & & \hspace{6cm}                    
   + ~ 6 \, \tau_6 \,  \bigg(  4 \, p^2 \, \frac{(pk) }{k^2} - \, (pk) \bigg)
   ~ + ~8 \,  \tau_8 \, \bigg(  p^2 - \frac{(pk)^2}{k^2}  \bigg)
\Bigg]
\nonumber \\
& &
\quad + ~  A\left(\left(p + \frac{k}{2}\right)^2\right) B\left(\left(p - \frac{k}{2}\right)^2\right) 
\Bigg[ 
   \, - \, 4 \,  \lambda_3 \, \bigg( \frac{ (pk)}{k^2} + 2 \, \frac{ p^2}{k^2}\bigg) 
   ~ + ~ 4 \, \tau_1 \,   \bigg(  p^2  - \frac{ (pk)^2}{k^2} \bigg)
   \nonumber \\
   & & \hspace{6cm}
  ~ + ~  2\, \tau_4 \,  \bigg(  p^2 k^2  + 2 \, p^2  (pk)   -  (pk)^2 - 2 \, \frac{(pk)^3}{k^2} \bigg)
~ + ~ 6 \, \tau_5 \,\bigg( 1 +  2 \, \frac{(pk)}{k^2}  \bigg)
   \nonumber \\
   & & \hspace{6cm}                    
   ~ + ~  4 \, \tau_7 \,  \bigg(   p^2 + 2 \, \frac{(pk)^2}{k^2} + 6 \, p^2 \, \frac{(pk)}{k^2}  \bigg)
\Bigg]
\nonumber \\
& &
\quad + ~  B\left(\left(p + \frac{k}{2}\right)^2\right) A\left(\left(p - \frac{k}{2}\right)^2\right)
\Bigg[
   4 \, \lambda_3 \,  \bigg(  \frac{  (pk) }{k^2} - 2 \, \frac{ p^2}{k^2} \bigg)
   ~ + ~ 4 \, \tau_1 \,  \bigg( p^2 - \frac{(pk)^2}{k^2}  \bigg)
 \nonumber \\
 & & \hspace{6cm}               
 + ~ 2  \, \tau_4 \, \bigg(  p^2 k^2   - 2 \, p^2 (pk) -   (pk)^2 + 2 \, \frac{(pk)^3}{k^2} \bigg)
 ~ + ~ 6 \, \tau_5 \,  \bigg( 1 - 2 \, \frac{(pk)}{k^2} \bigg)
 \nonumber \\
 & & \hspace{6cm}                 
 + ~ 4 \, \tau_7 \, \bigg(   p^2 + 2 \, \frac{(pk)^2 }{k^2} - 6 \, p^2 \, \frac{ (pk)}{k^2}   \bigg)
\Bigg]
\nonumber \\
& &
+B\left(\left(p + \frac{k}{2}\right)^2\right) B\left(\left(p - \frac{k}{2}\right)^2\right) 
\Bigg[
16 \, \lambda_1 \frac{1}{k^2} 
~ +~ 16 \, \lambda_2 \,   \frac{p^2}{k^2}
~ + ~ 8 \, \tau_2 \,  \bigg(  p^2 - \frac{(pk)^2}{k^2} \bigg)
 ~ + ~ 12 \, \tau_3 \, 
 ~ - ~ 24 \, \tau_6 \,  \frac{(pk)}{k^2} 
\Bigg] ~
\Bigg\} 
  \Bigg\}
\nonumber \\
& &
\nonumber \\
& &
= k^2 \, \left[  Z_3 \, 
  - i \, \frac{g^2}{3} \, Z_2 \, 
  \left( \Pi_0 (k^2)  + \frac{\Pi_1(k^2)}{k^2} \right) \right]
     \label{Eq:PhotonEqBasis}
\end{eqnarray}
where the form factors are 
\begin{equation}
\lambda_i = \lambda_i \big( (p-k/2)^2, \, (p + k/2)^2, \, k^2\big)
\qquad\mbox{ and }\qquad 
\tau_i = \tau_i \big( (p-k/2)^2, \, (p + k/2)^2, \, k^2\big) \ .
\label{TransFF-PhotonGapEq}
\end{equation}
The functions $\Pi_0 (k^2)$ and $\Pi_1(k^2)$ are regular finite integral expressions that can be written in terms of the fermion functions
$A$, $B$ and of the various photon-fermion vertex form factors $\lambda_i$ and $\tau_i$. 
These two function define the photon self-energy that is given by
\begin{equation}
\Pi(k^2) = \Pi_0 (k^2)  + \frac{\Pi_1(k^2) }{ k^2 } \ .
\end{equation}
From the renormalization conditions defined in Eq. (\ref{Renormalization-Conditions}), the constant associated with the photon renormalization is given in terms of the self-energy by
\begin{equation}
Z_3 ~ = ~ 1 ~ + ~  i ~ \frac{g^2}{3} ~ Z_2 ~ \Pi (\mu^2_B) \ .
\label{Eq:forZ3fromPi}
\end{equation}

From the definition of the photon self-energy, it follows that the function $\Pi_1(k^2)$ can, at zero photon momentum, lead to an infrared divergence in the photon self-energy. 
Indeed, for a $\mu_B = 0$, this case implies an infinite photon renormalization constant unless $Z_2$ cancels the would be divergence or $\Pi_1(0)$ vanish. 
The infrared divergences can be cancelled exactly if the later condition is fulfilled. 

In terms of the fermion propagator functions and vertex form factors, $\Pi_1(k^2)$ is given by
\begin{eqnarray}
&&
A \Big( (p + k/2)^2 \Big) \, A \Big( (p - k/2)^2 \Big) \Bigg\{
 - \, 8 \, \lambda_1 \,  p^2 
\, + \, 8 \, \lambda_2  \Big(   2 \,  p^4  \, - \,   (pk)^2\Big)
\nonumber \\
& &
\hspace{5.5cm}
+ ~
(pk) \bigg( 
-\, 8 \, \tau_2 \, p^2 \, (pk)
-\, 8 \, \tau_3 \, (pk)
+\, 24 \, \tau_6 \, p^2 \, 
-\, 8 \, \tau_8 \, (pk) \bigg)
\Bigg\}
\nonumber \\
&&
 + ~
A \Big( (p + k/2)^2 \Big) \, B \Big( (p - k/2)^2 \Big) \Bigg\{
 - \, 4 \, \lambda_3 \, \Big( 2 \,  p^2  \, + \, (pk)  \Big)
\nonumber \\
& &
\hspace{5.5cm}
+ ~(pk) \bigg( 
- \, 4 \, \tau_1 \, (pk)
-\,  4 \, \tau_4 \, (pk)^2
+\, 12 \, \tau_5 \, 
+\, 8 \, \tau_7 \, \Big(  \,  (pk) \, + \, 3 \, p^2 \,  \Big)
\bigg)
\Bigg\}
\nonumber \\
& &
+ ~ B \Big( (p + k/2)^2 \Big) \, A \Big( (p - k/2)^2 \Big) \Bigg\{
 4 \, \lambda_3  \Big( (pk)  \, - \, 2 \, p^2 \Big)
\nonumber \\
& &
\hspace{5.5cm}
+ ~ (pk) \bigg( 
- \, 4 \, \tau_1 \, (pk)
+\, 4 \, \tau_4 \, (pk)^2
- \, 12 \, \tau_5 \, 
-\, 8 \, \tau_7  \, \Big(  \, 3 \, p^2 \,  -\,  (pk) \Big)
\bigg)
\Bigg\}
\nonumber \\
& &
+ ~B \Big( (p + k/2)^2 \Big) \, B \Big( (p - k/2)^2 \Big) \Bigg\{
16 \, \lambda_1 \, 
\, + \, 16 \, \lambda_2 \,  p^2 
+~ (pk) \bigg( \, -\, 8 \, \tau_2 \, (pk) \, - \, 24 \, \tau_6 \, \bigg)
\Bigg\} \ ,
\label{Eq:Pi1k2}
\end{eqnarray}
where the factor
\begin{eqnarray}
- i \, \frac{g^2}{3} \, Z_2 \,  \int \frac{d^4 p}{(2 \, \pi)^4} ~  \frac{1}{A^2\left(\left(p + \frac{k}{2}\right)^2\right) \, (p + \frac{k}{2})^2 - B^2\left(\left(p + \frac{k}{2}\right)^2\right)} 
~                                                                             ~ \frac{1}{A^2\left(\left(p - \frac{k}{2}\right)^2\right) \, (p-\frac{k}{2})^2 - B^2\left(\left(p - \frac{k}{2}\right)^2\right)} 
\label{Eq:IntegrationFactor}
\end{eqnarray}
was omitted to simplify the notation. In Eq. (\ref{Eq:Pi1k2}) the contribution to $\Pi_1(k^2)$ due to the transverse form factors $\tau_i$ is proportional to
the scalar product $(pk)$. Then, if the $\tau_i$ are not singular at zero momentum, their contribution to $\Pi_1(k^2)$ vanish in the limit $k \rightarrow 0$,
and
at this kinematical point $\Pi_1(0)$ is a function of the longitudinal form factors $\lambda_i$ only. Indeed, at zero momentum, Eq. (\ref{Eq:Pi1k2}) becomes
\begin{eqnarray}
\widetilde{\Pi}_1(p^2) & = & 
8 \, \Bigg\{ 
A^2 \left( p^2 \right) \, p^2 \, \bigg[  - \,  \lambda_1 \, + \, 2 \, \lambda_2  \,  p^2 \bigg]
\, - \, 2 \, 
A \left( p^2 \right)\, B \left( p^2 \right) \,  \lambda_3 \, p^2 
\, + \,
2 \, B^2  \left( p^2 \right) \, \bigg[  \, \lambda_1 \,  \, + \,  \lambda_2 \,  p^2  \bigg] \Bigg\} \ ,
\label{Eq:Pi1k20}
\end{eqnarray}
where $\lambda_i = \lambda_i \big( p^2, \, p^2, \, 0\big)$ and Eq. (\ref{LongVertex-ZeroMom}) can be used to write the form factors in terms of the functions $A$,
$B$  and their derivatives evaluated at $p^2$. From the above considerations it follows that the function $\Pi_1 (k^2)$ is a Lorentz scalar smooth function, 
for all the momenta, that, for small photon momentum, can be written as
\begin{equation}
\Pi_1 (k^2) = \Pi^\prime_1(0) \, k^2 + \mathcal{O}(k^4) \ ,
\end{equation}
where $\Pi^\prime_1$ is the derivative of $\Pi_1$ with respect to $k^2$.

If all the form factors $\lambda_i$ and $\tau_i$ are smooth functions at low momentum, then the DSE for $1/D(0)$ requires only the longitudinal components of
$\Gamma^\mu$, i.e. the contribution of the $\lambda_i$'s. The integral equation reads
\begin{eqnarray}
\frac{1}{D(0)} = - i \, \frac{g^2}{3} \, Z_2 \,  \Pi_1(p^2)  = 
- i \, \frac{g^2}{3} \, Z_2\,  \int \frac{d^4 p}{(2 \, \pi)^4} ~  \frac{\widetilde{\Pi}_1(p^2) }{\Big[ A^2\left(p^2\right) \, p^2 - B^2\left(p^2\right)\Big]^2} \ .
\label{Eq:DSEPhotonZero0}
\end{eqnarray}
The above reasoning shows that the masslesness of the photon and the infrared divergences in the photon self-energy are related. The absence of the IR
divergences in the photon self-energy is equivalent to state that the photon is a massless particle.
This statement is independent of the value of the QED coupling constant and, therefore, it should be true even beyond the domain of applicability of perturbation theory.

In terms of the photon self-energy, the photon propagator reads
\begin{equation}
D(k^2) = \frac{1}{Z_3\, \, k^2 \,\, \left( 1 \,   - i \, \frac{g^2}{3} \, \frac{Z_2}{Z_3} \, \Pi (k^2)   \right) } \ .
\end{equation}
If the self-energy is finite for all $k$, then the propagator divergences at $k^2 = 0$. More, at low $k^2$, the photon propagator reproduces the 
prediction of tree-level perturbation theory.
The masslesness of photon is responsible for having an IR safe self-energy and it implies also that at low momentum $D(k^2)$ diverge as $1/k^2$.
Note, however, that this does prevent $D(k^2)$ from having poles at $k^2 \ne 0$. Indeed, a solution of
\begin{equation}
1 \,   - i \, \frac{g^2}{3} \, \frac{Z_2}{Z_3} \, \Pi (k^2_0)  = 0 \ ,
\end{equation}
implies a pole in the propagator at $k^2_0$. 
Perturbation theory looks for a solution of the integral equation that is a ``small'' correction to the lowest order solution. 
Then, there is always a range of values for the coupling constant where the above equation does not have a solution and, within this range of values for $g$,  
$D(k^2)$ has a simple pole at $k^2 = 0$ and no further singularities. However,
the presence of singularities in $D(k^2)$ for $k^2 \ne 0$ and for larger coupling constants is not excluded.

The above conclusions do not hold if any of the transverse form factors is singular at zero momentum. 
In this case, the possible singular form factor(s) contributing to $1/D(0)$ can give rise to a mass-like term. This is the gauge invariant Schwinger mechanics for
mass generation that seems to take place in QCD, making the gluon propagator finite and non-vanishing
at zero momentum. Note, however, that our conclusion does not hold for QCD as the gluon DSE include terms requiring three-gluon, four-gluon and 
ghost-gluon vertices that change the details of the integral equation.

The conventional perturbative result sets, to lowest order in the coupling constant, 
\begin{eqnarray}
\widetilde{\Pi}_1(p^2) ~ = ~  - \, 8 \,  \,  p^2  \, + \, 16 \,  m^2
\label{Eq:Pi1k20Pert}
\end{eqnarray}
and, therefore,
\begin{eqnarray}
\frac{1}{D(0)} = 
 i \, \frac{8 \, g^2}{3} \, Z_2 \,  \int \frac{d^4 p}{(2 \, \pi)^4} ~  \left( \frac{1}{ p^2 - m^2 }  - \frac{m }{\left(  \, p^2 - m^2 \, \right)^2}  \right)\ .
\label{Eq:DSEPhotonZero0Pert}
\end{eqnarray}
In this case, the photon appears as a massive boson with a mass that is proportional to the fermion mass $m$, see the discussion in 
\cite{Oliveira:2025ftr}, unless $Z_2 = Z_1 = 0$. 
This last condition implies a vanishing physical fermion field and QED would become a trivial free field boson-like theory.

\section{The fermion gap equation \label{Sec:fermions}}

The fermion gap equation is an infrared safe equation, and is well behaved in the low momentum limit. In this integral equation there is a term that multiplies the photon propagator,
and another term that is multiplied by $\xi / k^2$, that comes from the longitudinal component of the bosonic propagator. Recall that at low momentum
the photon propagator goes as $1/k^2$. The two terms are multiplied by $k^3$, that comes from the integral measure, and at low momentum the
leading term in the loop momentum is linear in $k$, making the fermion gap equation infrared safe.

Another interesting case that can be studied is the chiral fermionic theory that is defined by having $B \left( p^2 \right) = 0$ for any momentum. 
This definition implies that $\lambda_3 = 0$ and if it holds, then
\begin{eqnarray}
\widetilde{\Pi}_1(p^2) & = &  8 \, 
A^2 \left( p^2 \right) \, p^2 \, \bigg[  - \,  \lambda_1 \, + \, 2 \, \lambda_2  \,  p^2 \bigg] ~ = ~
A^2 \left( p^2 \right) \, p^2 \, \left[  - \,  A \left( p^2 \right) \, + \, \frac{d A \left( p^2 \right)}{d p^2}  \,  p^2 \right] 
\ .
\label{Eq:Pi1k20Chiral}
\end{eqnarray}
A massless photon requires that
\begin{equation}
- \,  A \left( p^2 \right) \, + \, \frac{d A \left( p^2 \right)}{d p^2}  \,  p^2 = 0 \ ,
\label{PhotonChiral}
\end{equation}
whose solution is
\begin{equation}
  A \left( p^2 \right) = C \, p^2 \ ,
\end{equation}
where $C$ is a constant of integration. It follows that the propagator for a chiral fermion reads
\begin{equation}
  S(p) = \frac{\slashed{p}}{C \, p^4} \ .
\end{equation}
The requirement of having a massless photon in QED translates into a chiral fermion propagator with an improved ultraviolet behavior,
in the sense that, at large momentum, the chiral propagator goes as $1/p^3$, to be compared with the conventional perturbation theory where the
propagator scales as $1/p$  in the ultraviolet regime. The implications of a chiral theory to the gap equation and to the transverse form factors
are further explored in App. \ref{Sec:ChiralGapEq}.

Although we have explored how Eq. (\ref{PhotonChiral}) impacts in chiral theories with a massless photon, the definition of the chiral theory 
is an integral relation involving $\lambda_1$ and $\lambda_2$ and, therefore, deviations from the solution just considered are not forbidden. In particular, one
should recall that the simplest solution of the vertex WTI to set $\lambda_1$ and $\lambda_2$, see Eqs (\ref{EQ:L1}) to (\ref{EQ:L4}), 
was used but the WTI for the vertex does not give these form factors in an unambiguous way.
Moreover, it is well known that for sufficiently large couplings dynamical chiral symmetry breaking takes place, see e.g. 
\cite{Johnson:1973pd,Curtis:1990zr,Roberts:1994dr,Albino:2018ncl,Lessa:2022wqc,Oliveira:2024tne} and references therein,
and, for sufficiently large couplings, fermions become massive and there are no chiral fermions in QED. For QED, the value of the coupling where
dynamical chiral symmetry breaking takes places is gauge dependent and increase with $\xi$.
The above reasoning for chiral fermions apply only to relatively small values of the coupling constant.

\section{The vertex equation \label{Sec:VertEq}}

Let us discuss infrared divergences within the photon-fermion vertex DSE. For a vanishing photon momentum the integral equation reads
\begin{eqnarray}
& & 
   {\Gamma}^\mu (p, \, -p; \, 0)  ~  =  ~  Z_2 \, \gamma^\mu  ~ + ~ i \, g^2 \,  Z_2 \, \int \frac{d^4q}{(2 \, \pi)^4} ~D_{\zeta\zeta^\prime}(q) \nonumber \\
   & & \hspace{0.5cm}
    \Bigg\{
                                \gamma^\zeta ~ S(p-q) ~ {\Gamma}^\mu ( p - q, \, -p + q; \, 0) ~
                                S(p-q) ~ {\Gamma}^{\zeta^\prime} (p -q, \, -p; q) 
                 ~ + ~
                                \gamma^\zeta \, S(p-q) \, {\Gamma}^{\zeta^\prime\mu} (p - q, \, -p; \, q , \, 0) 
                                \Bigg\} 
\nonumber \\
& & 
~  =  ~  Z_2 \, \gamma^\mu  ~ - ~ i \, g^2 \,  Z_2 \, \int \frac{d^4q}{(2 \, \pi)^4} ~D(q^2) \nonumber \\
   & & \hspace{1.5cm}
    \Bigg\{
                                \gamma_\zeta ~ S(p-q) ~ {\Gamma}^\mu ( p - q, \, -p + q; \, 0) ~
                                S(p-q) ~ {\Gamma}^{\zeta} (p -q, \, -p; q) 
                 ~ + ~
                                \gamma_\zeta \, S(p-q) \, {\Gamma}^{\zeta\mu} (p - q, \, -p; \, q , \, 0) 
                                \Bigg\}  
                                \nonumber \\
& &  
~- ~ i \, g^2 \,  Z_2 \, \int \frac{d^4q}{(2 \, \pi)^4} ~ \left( \frac{\xi}{q^2} - D(q^2) \right) ~ \frac{1}{q^2} 
\nonumber \\
& & \hspace{1.6cm}
    \Bigg\{ \slashed{q} ~ S(p-q) ~ {\Gamma}^\mu ( p - q, \, -p + q; \, 0) ~ S(p-q) ~ S^{-1}(p)
                      ~ - ~\slashed{q} ~ S(p-q) ~ {\Gamma}^\mu ( p , \, -p ; \, 0)  \Bigg\}    
                                \, ,
                                \label{DSE-vertexzero}
\end{eqnarray}
where in writing the last form of the equation the definition of the photon propagator, see Eq. (\ref{Def:PhotonProp}), was taken into account,
together with the vertex WTI
\begin{equation}
k_\mu \, \Gamma^\mu(p, \, -p-k ; \, k)= S^{-1}(p + k) ~ - ~  S^{-1}(p) , 
\end{equation}
and the two-photon-two-fermion vertex WTI
\begin{equation}
k_\mu \, \Gamma^{\mu\nu} (p, \, -p-k-q; \, k, \, q) ~ = ~ \Gamma^\nu(p, \,  -p-q; \, q) ~ - ~ \Gamma^\nu (p + k, \, - p-k-q; \, q)  \ .
\end{equation}
For a derivation of the WTI see e.g. \cite{Eichmann:2012mp,Oliveira:2022bar} and references therein. 
To understand the infrared properties in the vertex equation, the low momenta limit of the  integrand in Eq. (\ref{DSE-vertexzero}) is considered.
In the limit $q \rightarrow 0$, the last term within brackets vanishes exactly and it is infrared safe. 
The term within brackets in the second term of the equation has two contributions. The first is independent of $\Gamma^{\mu\nu}$ and as long as
$D(q^2) \sim 1/(q^2)^{1 + \iota}$ with $\iota > -1/2$, as occurs for the tree-level photon propagator that has $\iota = 0$, the term is also infrared safe.
Possible infrared divergences can only be associated with the two-photon-two-fermion one-particle irreducible Green function 
$\Gamma^{\mu\nu}$. In perturbation theory $\Gamma^{\mu\nu}$ is proportional to $g^4$ and, to lowest order in the coupling constant
$\Gamma^{\mu\nu}$ is infrared safe. Further, the DSE equation for $\Gamma^{\mu\nu}$, see \cite{Oliveira:2022bar}, suggests that, at least for massive fermions, 
$\Gamma^{\mu\nu}$ gives also an infrared safe contribution to the vertex. 
Our conclusion being that, in QED with massive fermions, 
there are no infrared singularities associated with the photon-fermion vertex.

\section{Schwinger mechanism: the case of singular transverse form factors \label{Sec:Schwinger}}

In order to investigate possible singular behavior for the transverse form factors let us set $\tau_i = \widetilde{\tau}_i / k^2$, where $k$ is the photon momentum,
and assume that the $\widetilde{\tau}_i$ are smooth functions of the momenta, i.e. that they are not singular at $k = 0$. 
Another possibility of having a singular behavior, that will not be pursued here, it to have $\tau_i = \widetilde{\tau}_i / (pk)$ with a regular $\widetilde{\tau}_i$ at the origin.
From the point of view of the fermion gap equation, such a redefinition of the transverse form factor does not compromise any of the $\tau_i$.
Indeed, in the fermion gap equation, they are multiplied by powers of $k$ that prevent the appearance of infrared divergences, see e.g. the appendices in \cite{Oliveira:2025ftr}.

The writing of the transverse form factors as defined now generate, in the photon self-energy, terms that go with the photon momentum up to $1/k^4$. 
Looking to the contributions of the $\tau_i$'s only, they read
\begin{eqnarray}
&&
A \Big( (p + k/2)^2 \Big) \, A \Big( (p - k/2)^2 \Big) \, \Bigg\{
 2 \, \widetilde{\tau}_2 \, \left( p^2 \, + \, 4  \, \frac{p^4}{k^2}   \,  - \, \frac{(pk)^2}{k^2} \, - \, 4 \, p^2 \, \frac{ \, (pk)^2}{\, k^4 \,} \right)
+\, \widetilde{\tau}_3 \, \left(3 \, - \, 4 \, \frac{p^2}{k^2} \, - \, 8 \, \frac{ \, (pk)^2}{\, k^4 \,}\right)
\nonumber \\
& &
\hspace{6cm}
+\, 6 \, \widetilde{\tau}_6  \, \left( -  \, \frac{(pk)}{k^2} \,  + \, 4 \, p^2 \, \frac{ \, (pk) \,}{\, k^4 \,} \right)
+\, 8 \, \widetilde{\tau}_8  \, \left( \, \frac{p^2}{k^2} \,- \, \frac{\, (pk)^2}{\, k^4 \,}\right)
\Bigg\}
\nonumber \\
& &
+ ~ A \Big( (p + k/2)^2 \Big) \, B \Big( (p - k/2)^2 \Big) \, \Bigg\{
 4 \, \widetilde{\tau}_1  \, \left( \, \frac{p^2}{k^2} \,- \, \frac{ \, (pk)^2}{\, k^4 \,}\right)
+\, 2 \, \widetilde{\tau}_4  \, \left(   \, p^2 \,  + \, 2 \, p^2 \, \, \frac{ (pk) }{k^2}  \,- \, \frac{(pk)^2}{k^2} \,- \, 2 \, \frac{ \, (pk)^3}{\, k^4 \,} \right)
\nonumber \\
& &
\hspace{6cm} 
+\, 6 \, \widetilde{\tau}_5  \, \left( \frac{1}{k^2} \, + \, 2 \, \frac{\, (pk) \,}{\, k^4 \,} \right)
+\, 4 \, \widetilde{\tau}_7 \, \left( \, \frac{p^2}{k^2} \,  + \, 2 \, \frac{\, (pk)^2}{\, k^4 \,} \, + \, 6 \, p^2 \, \frac{ \, (pk) \,}{\, k^4 \,}  \right)
\Bigg\}
\nonumber \\
& & 
+ ~  B \Big( (p + k/2)^2 \Big) \, A \Big( (p - k/2)^2 \Big) \, \Bigg\{
 4 \, \widetilde{\tau}_1 \, \left( \, \frac{p^2}{k^2} \,- \, \frac{\, (pk)^2}{\, k^4 \,}\right)
+\, 2 \, \widetilde{\tau}_4  \, \left( \, p^2 \, - \, \frac{(pk)^2}{k^2} \, - \, 2 \, p^2 \, \, \frac{(pk)}{k^2}\, + \, \frac{2 \, (pk) \,^3}{\, k^4 \,}  \right)
\nonumber \\
& &
\hspace{6cm}
+\, 6 \, \widetilde{\tau}_5  \, \left( \frac{1}{k^2} \, - \, 2 \, \frac{\, (pk) \,}{\, k^4 \,}\right)
+\, 4 \, \tau_7  \, \left(  \, \frac{p^2}{k^2} \,- \, 6 \, p^2 \,\frac{ \, (pk) \,}{\, k^4 \,} \, +\, 2 \, \frac{ \, (pk)^2}{\, k^4 \,} \right)
\Bigg\}
\nonumber \\
& & 
+ ~ B \Big( (p + k/2)^2 \Big) \, B \Big( (p - k/2)^2 \Big) \, \Bigg\{
 8 \, \widetilde{\tau}_2 \, \left( \, \frac{p^2}{k^2} \,- \, \frac{ \, (pk)^2}{\, k^4 \,}\right)
\, +12 \, \widetilde{\tau}_3 \, \frac{1}{k^2} 
\, - \, 24 \, \widetilde{\tau}_6 \,  \frac{\, (pk) \,}{\, k^4 \,}
\Bigg\}
\nonumber \\
& &
= ~ \widetilde{\Pi}_0 (p^2, k^2, (pk)) + \frac{\widetilde{\Pi}_1 (p^2, k^2, (pk))  }{k^2} + \frac{\widetilde{\Pi}_2(p^2, k^2, (pk))  }{k^4}
\end{eqnarray}
where the notation is inspired in that used before, see Eq. (\ref{Eq:PhotonEqBasis}) and subsequent equations,
and the factor (\ref{Eq:IntegrationFactor}) was not written explicitly to simplify the
notation. The momentum dependence of the $\widetilde{\tau}_i$'s is as in Eq. (\ref{TransFF-PhotonGapEq}).
We recall the reader that the longitudinal form factors contribute, at most, with a $1/k^2$ term to the photon self-energy that mixes
with the contribution of $\widetilde{\Pi}_1$. It follows that
\begin{eqnarray}
&&
\widetilde{\Pi}_0 (p^2, k^2, (pk)) = A \Big( (p + k/2)^2 \Big) \, A \Big( (p - k/2)^2 \Big) \, \Bigg(
 2 \, \widetilde{\tau}_2 \, p^2 
+\, 3 \, \widetilde{\tau}_3 \, \Bigg)
\nonumber \\
& &
\qquad\qquad
 + ~  \Bigg( A \Big( (p + k/2)^2 \Big) \, B \Big( (p - k/2)^2 \Big) 
+ ~  B \Big( (p + k/2)^2 \Big) \, A \Big( (p - k/2)^2 \Big) \Bigg)  
\, 2 \, \widetilde{\tau}_4  \,  p^2
\label{SchwingerIRsafe}
\end{eqnarray}
is an infrared safe contribution to the self-energy, while
\begin{eqnarray}
&&
\widetilde{\Pi}_1 (p^2, k^2, (pk)) = A \Big( (p + k/2)^2 \Big) \, A \Big( (p - k/2)^2 \Big) \, 2 \, \Bigg[
  \widetilde{\tau}_2 \, \Big(  \, 4  \, p^4   \,  - \, (pk)^2  \Big)
-\, 2 \, \widetilde{\tau}_3 \, p^2
-\, 3 \, \widetilde{\tau}_6  \, (pk)
+\, 4 \, \widetilde{\tau}_8  \, p^2
\Bigg]
\nonumber \\
& &
+ ~ A \Big( (p + k/2)^2 \Big) \, B \Big( (p - k/2)^2 \Big) \, 2 \, \Bigg[
 2 \, \widetilde{\tau}_1  \, p^2
+\,  \widetilde{\tau}_4  \, (pk) \, \Big(    \, 2 \, p^2    \,- \, (pk) \Big)
+\, 3 \, \widetilde{\tau}_5  \, 
+\, 2 \, \widetilde{\tau}_7 \, p^2
\Bigg]
\nonumber \\
& & 
+ ~  B \Big( (p + k/2)^2 \Big) \, A \Big( (p - k/2)^2 \Big) \, 2 \, \Bigg[
 2 \, \widetilde{\tau}_1 \,  p^2
-\,  \widetilde{\tau}_4  \, (pk) \, \Big(  \, 2 \, p^2 \,  + \, (pk)   \Big)
+\, 3 \, \widetilde{\tau}_5  \, 
+\, 2 \, \widetilde{\tau}_7  \, p^2
\Bigg]
\nonumber \\
& & 
+ ~ B \Big( (p + k/2)^2 \Big) \, B \Big( (p - k/2)^2 \Big) \, 4 \, \Bigg[
 2 \, \widetilde{\tau}_2 \,  p^2
\, +\, 3 \, \widetilde{\tau}_3 \, 
\Bigg]
\end{eqnarray}
and
\begin{eqnarray}
&&
\widetilde{\Pi}_2 (p^2, k^2, (pk)) = A \Big( (p + k/2)^2 \Big) \, A \Big( (p - k/2)^2 \Big) \, 8 \, (pk) \Bigg[
 - \,  \widetilde{\tau}_2 \,  p^2 \, (pk)
-\,  \widetilde{\tau}_3 \,   (pk)
+\, 3 \, \widetilde{\tau}_6  \,  p^2 
-\,  \widetilde{\tau}_8  \,   (pk)
\Bigg]
\nonumber \\
& &
+ ~ A \Big( (p + k/2)^2 \Big) \, B \Big( (p - k/2)^2 \Big) \, 4 \, (pk) \, \Bigg[
 - \,  \widetilde{\tau}_1  \,  (pk)
-\,  \widetilde{\tau}_4  \,   (pk)^2
+\, 3 \, \widetilde{\tau}_5   
+\, 2 \, \widetilde{\tau}_7 \, \Big( \,  (pk) \, + \, 3 \, p^2 \,  \Big)
\Bigg]
\nonumber \\
& & 
+ ~  B \Big( (p + k/2)^2 \Big) \, A \Big( (p - k/2)^2 \Big) \, 4 \, (pk) \, \Bigg[
 - \,  \widetilde{\tau}_1 \,  (pk)
+\,  \widetilde{\tau}_4  \,  (pk)^2
-\, 3 \, \widetilde{\tau}_5   
+\, 2 \, \widetilde{\tau}_7  \, \Big(   \, (pk) \,- \, 3 \, p^2 \,  \Big)
\Bigg]
\nonumber \\
& & 
+ ~ B \Big( (p + k/2)^2 \Big) \, B \Big( (p - k/2)^2 \Big) \, 8 \, (pk) \, \Bigg[
 - \,  \widetilde{\tau}_2 \,  (pk)
\, - \, 3 \, \widetilde{\tau}_6 \, 
\Bigg]
\end{eqnarray}
are associated with the $1/k^2$ and $1/k^4$ terms, respectively. 
These two terms can lead to infrared divergent contributions to the photon self-energy. 
In the limit of zero photon momentum
\begin{eqnarray}
& & 
\widetilde{\Pi}_1 (p^2, \, 0, \, 0) =
A^2 \big( p^2 \big) \,  \, \Bigg\{
 8 \, \widetilde{\tau}_2 \, \, p^4
-\, 4 \, \widetilde{\tau}_3 \, p^2
+\, 8 \, \widetilde{\tau}_8  \, \, p^2
\Bigg\}
+ ~ 2 \, A \big( p^2 \big) \, B \big( p^2 \big) \, \Bigg\{
 4 \, \widetilde{\tau}_1  \, p^2
+\, 6 \, \widetilde{\tau}_5  
+\, 4 \, \widetilde{\tau}_7 \, \, p^2
\Bigg\}
\nonumber \\
& & 
\hspace{4cm}
+ ~ B^2 \big( p^2 \big) \,  \Bigg\{
 8 \, \widetilde{\tau}_2 \,  \, p^2
\, +12 \, \widetilde{\tau}_3 \, 
\Bigg\} \ ,
\label{Eq:widetildePi1}
\end{eqnarray}
while $\widetilde{\Pi}_2 (p^2, k^2, (pk))$ being proportional to $(pk)$ implies that
\begin{equation}
\widetilde{\Pi}_2 (p^2, \, 0, \, 0) = 0 \ .
\end{equation}
It turns out that in the photon DSE the term associated with $\Pi_1$ can be made infrared finite as before. The term ${\Pi}_2$ seems to be infrared safe but
it can give a contribution that, being  proportional to derivatives of the $\widetilde{\tau}_i$ and/or $A$ and $B$, looks like a photon mass term.

In the photon gap equation, the self-energy is multiplied by $k^2$ and, therefore, possible contributions that can generate a photon mass term can only
come from $\widetilde{\Pi}_1$ or $\widetilde{\Pi}_2$. Given that $\widetilde{\Pi}_2 (p^2, \, 0, \, 0) = 0$, any contribution coming this function has
to call for derivatives of $\widetilde{\tau}_i$ and/or of $A$ and $B$. Let us ignore this type of terms and consider only possible non-derivatives contributions.
Then, the DSE for the photon propagator reads
\begin{equation}
\frac{1}{D(0)}  ~  =  ~
- i \, \frac{g^2}{3} \, Z_2\,  \int \frac{d^4 p}{(2 \, \pi)^4} ~  \frac{\widetilde{\Pi}_1(p^2, \, 0, \, 0) }{\Big[ A^2\left(p^2\right) \, p^2 - B^2\left(p^2\right)\Big]^2}  \ ,
\label{Eq:PhotonMassSchwinger}
\end{equation}
where $\widetilde{\Pi}_1 (p^2, 0, \, 0)$ is given in Eq. (\ref{Eq:widetildePi1}). This contribution looks like a photon mass term that is associated with
a singular behavior from the transverse form factors $\tau_{1,2,3,5,7,8}$. On the other hand, the transverse form factors $\tau_{4,6}$ are not able to 
generate a photon mass via the Schwinger mechanism. Note that to avoid possible UV divergencies in Eq. (\ref{Eq:PhotonMassSchwinger}), the 
integration over the loop momentum constrains the possible UV behavior of all the $\widetilde{\tau}_i (p^2, \, p^2, \, 0)$ that are present in the equation. 

The complete mass term that appears in the photon gap equation, i.e. the $1/k^2$ term in the self-energy, includes the contribution of both the longitudinal, 
see Eq. (\ref{Eq:Pi1k20}), and the transverse, see Eq. (\ref{Eq:widetildePi1}), form factors to $\widetilde{\Pi}_1$. 
As discussed, $\widetilde{\Pi}_2$ can also contribute with a term that is proportional to the derivatives of the $\widetilde{\tau}_i$.
Again, the requirement of having a finite photon self-energy implies, once more,
a massless photon, modulo possible derivatives contributions. This requirement provides also an integral relation between the longitudinal and
transverse form factors. 
Within the Schwinger mechanism, the link between the mass of the photon and an infrared safe photon self-energy is not as clear
as when assumes a smooth behavior of the transverse form factors. Note, however, that the condition of having a vanishing photon mass can be recovered 
by requiring that the numerator in the $1/k^2$ vanishes exactly. Indeed, despite that formally in $\Pi (k^2)$ there are terms up to $1/k^4$,
given that $\widetilde{\Pi}_2(0) = 0$ then, at low $k$ only the $1/k^2$ survive; one is back to the case discussed in Sec. \ref{Sec:PhotonGapEq}.

As discussed in \cite{Oliveira:2025ftr}, the form factors $\tau_i$'s are related to traces of the photon-fermion vertex with various types of Dirac
matrices, a connection that needs to be further explored to understand if solutions of the vertex DSE as considered here are possible or not. If not, then
the origin of the singularities in the Schwinger mechanism has to be looked outside QED.

\section{Summary and Conclusion \label{Sec:Conclusions}}

In the current work the infrared properties of QED are investigated within the framework provided by the Dyson-Schwinger equations. Of all the DSE
only the propagators and the vertex equations are considered. As discussed, without computing the solution of the integral equations, we are 
able to derive a number of results valid for QED and for any value of the coupling constant. From the photon gap equation, demanding that the photon
self-energy is finite for any momentum and assuming a vertex that is always a smooth function of the momenta, the absence of infrared divergences and
the masslessness of the photon are equivalent statements. Moreover, it implies that the photon propagator at low momenta behave as its
tree-level perturbative solution

For small values of the coupling constant, the analysis of the DSE suggests that the photon propagator has a unique simple pole at $k^2=0$. For sufficiently
small $g$ no further singularities are allowed in $D(k^2)$.
However, for sufficiently large couplings the possibility that $D(k^2)$ has singularities away from the origin is not excluded. The DSE gives no 
information about these extra possible poles, neither on its nature.
From the fermion gap equation, it is well known that in QED and for sufficiently large couplings, dynamical chiral symmetry breaking takes place.
A mass generation mechanism for the photon at sufficiently large couplings was not yet observed but, given the nature of the solutions of the fermion propagator
DSE for large values of the coupling constant, it would not be a complete surprise if, for large $g$, the photon would become massive.

Our investigation of the fermion gap equation and of the vertex equation at low momenta show that both are infrared safe, 
i.e. no divergences seem to appear associated with these equations at low momenta.
The  analysis herein suggests that QED is an infrared free theory and that the infrared singularities observed in standard perturbation theory are related to the
perturbative expansion itself, that is problematic in handling massless particles. 
In perturbation theory the IR divergences require a proper treatment and a number of alternative approaches to solve the IR problem can be found in the literature.

Besides conventional QED, the gauge invariant Schwinger mechanism for mass generation is also discussed assuming that the transverse form factors $\tau_i$
are singular for a vanishing photon momentum $k$. The analysis performed assume that they all behave as $1/k^2$, for small $k$. 
In this case, a contribution for the photon self-energy coming from the $\tau_i$ mixes with a contribution associated with the longitudinal form factors $\lambda_i$
and, proceeding as before, a finite self-energy it is not enough to ensure that the photon keeps being a massless particle. Indeed, there is a non-trivial contribution to the photon
self-energy, that is associated with derivatives of the form factors and fermion propagator functions, that can result in a photon mass-like term in the photon propagator.
On the other hand, assuming that the singular $\tau_i$ appear as mass-like contribution in the propagator, we are able to identify those operators which can
generate a mass term. Our findings, show that all the transverse operators but $T^{(5)}$ and $T^{(6)}$ can generate a photon mass. 
The corresponding form factors associated with these two operators appear in the fermion gap equation and, independently, could be at the origin of dynamical 
chiral symmetry breaking. 

In what concerns the Schwinger mechanism, it remains to explain the possible origin of the singularities in the transverse form factors. Indeed, we do not
know if the photon-fermion vertex DSE allow for such type of solutions. The general solution of this equation, see \cite{Oliveira:2025ftr},
 favors an overall factor of $1/(p^2 k^2 - (pk)^2)$ in the transverse form factors that, in some cases, a closer look to the details to the solutions kills explicitly this 
overall contribution.
It is not clear if such overall factor survives in the vertex solution or is simply an intermediate artefact due to the way the solutions are written.

The discussion in the current work considers QED in the Minkowski spacetime but, in principle, it also applies to the Euclidean formulation of theory. 
Moreover, some of the results derived can be extended to the case of complex momentum. An example is  the number and nature of the photon propagator 
poles at complex momenta for small values of the coupling constant. 
The case of a propagator with complex conjugate poles is not being considered within QED, but it has raised various
discussions for the gluon propagator, where the complex conjugate poles are, for some authors, identified as a sign of gluon confinement. 
The discussion on complex poles for the gluon has even motivated the formulation of an axiomatic to include for that possibility in QFT.
Of course, in order to have the full picture, even for QED, we would like to be able to build the solution of the DSE. Unfortunately, this does
not seem to be an easy task, even for the case of the Abelian gauge theory. However, the results derived can be tested with the help of the
Euclidean formulation of the theory on a finite lattice, i.e. with Monte Carlo lattice QED simulations.

\section*{Acknowledgements}

This work was financed through national funds by FCT - Funda\c{c}\~ao para a Ci\^encia e a Tecnologia, I.P. in the framework of the projects UIDB/04564/2020 and UIDP/04564/2020, with DOI identifiers 10.54499/UIDB/04564/2020 and 10.54499/UIDP/04564/2020, respectively.
This work was financed through national funds by FCT - Fundação para a Ciência e Tecnologia, I.P. in the framework of the project UID/04564/2025.
We would like to thank Tobias Frederico for helpful discussions.

\begin{appendix}

\section{The gap equation for chiral fermions \label{Sec:ChiralGapEq}}

For completeness, herein the DSE for the fermion propagator is given in terms of the longitudinal and transverse form factors.
For $B(p^2) = m = 0$, that implies a $A(p^2) = C \, p^2$, the scalar component of the gap equation can be written as
\begin{eqnarray}
& & 
  0  ~ = ~ \int \frac{d^4 k}{( 2 \, \pi )^4} ~  \frac{D(k^2)}{A((p-k)^2) \,(p-k)^2 }  
     \nonumber \\
     & & \hspace{1.5cm}
 \Bigg\{ ~
               \tau_1 \, k^2 \, p^2 \bigg( 1 - \frac{(pk)^2}{k^2 \, p^2} \bigg)
              + ~ \tau_4 \, k^4 \, p^2 \bigg( 1 -  \frac{ (pk) }{k^2} -  \frac{(pk)^2}{p^2 \, k^2}  + \frac{(pk)^3}{p^2 \, k^4}  \bigg)
      \nonumber \\
      & & \hspace{2.5cm}\qquad
              ~ + ~ 3 \, \tau_5 \, k^2 \bigg( 1 - \frac{(pk)}{k^2} \bigg)
          ~  + ~   \tau_7 \, k^4 \bigg( \frac{3 }{2}  + 4 \,  \frac{p^2}{k^2}  - \frac{15 }{2} \, \frac{ (pk) }{k^2} - 6 \, \frac{p^2}{k^2} \, \frac{(pk)}{k^2} + 8 \,  \frac{(pk)^2}{k^4} \bigg)
 ~
\Bigg\}     \ .
      \label{DSE-Fermion-scalar}
\end{eqnarray}
On the other hand, the vector component of the same equation reads
\begin{eqnarray}
& & 
 A(p^2)  = 1
     + \, i \, g^2 \, \int \frac{d^4 k}{( 2 \, \pi )^4} ~   \frac{1}{A((p-k)^2) \,(p-k)^2 }\Bigg\{
\nonumber \\
& & \qquad
D(k^2) ~
     \Bigg[ ~
         \lambda_1 \,  \bigg( - 1 + 3 \, \frac{(pk)}{p^2} - \, 2 \, \frac{(pk)^2}{p^2 \, k^2}\bigg)
        ~ + ~ 2 \, \lambda_2 \, k^2  \bigg( 1 + 2 \, \frac{p^2}{k^2}  - 2 \, \frac{ (pk) }{k^2} -  \frac{(pk)^2}{p^2 \, k^2}
                                                                      - \, 2 \,  \frac{ (pk)^2}{k^4} + 2 \, \frac{(pk)^3}{p^2 \, k^4} \bigg)
         \nonumber \\
        & &     \hspace{1.5cm}\qquad
         + ~ \tau_2 \, k^4  \bigg( 1 + 2  \, \frac{p^2}{k^2}  - 2 \,  \frac{(pk)}{k^2}  - \frac{(pk)^2}{p^2 \, k^2}  - 2 \,  \frac{ (pk)^2}{k^4} + 2  \, \frac{(pk)^3}{p^2 \, k^4} \bigg)
        + ~ \tau_3 \, k^2  \bigg( - \, 1 + 3 \, \frac{(pk)}{p^2} - 2 \, \frac{(pk)^2}{p^2 \, k^2} \bigg)
        \nonumber \\
      & & \hspace{2.cm}  \qquad
        + ~ 3 \, \tau_6 \, k^2 \bigg(  1 -  \frac{(pk)}{p^2} - 2 \frac{(pk)}{k^2} + 2 \frac{(pk)^2}{p^2 \, k^2} \bigg)
        ~ + ~ 2\, \tau_8 \, k^2  \bigg( 1 -  \frac{(pk)^2}{p^2 \, k^2} \bigg)
                     ~~\Bigg]
 \nonumber \\
 & & \qquad
 + ~ \frac{\xi}{k^2}
     \Bigg[
               \lambda_1 \,  \bigg( - 1 \, - \frac{(pk)}{p^2}  + 2 \, \frac{ (pk)^2}{p^2 \, k^2} \bigg)
    ~           + ~ \lambda_2 \, k^2 \bigg( 1 - \frac{(pk)}{p^2} - 4 \, \frac{(pk)}{k^2}  + 4 \,  \frac{(pk)^2}{p^2 \, k^2}  + 4 \, \frac{(pk)^2}{k^4} - 4 \, \frac{(pk)^3}{p^2 \, k^4} \bigg)
               \Bigg]
~ \Bigg\} \ .
     \label{DSE-Fermion-vector}
\end{eqnarray}
In both cases, the form factors are short notations for
\begin{equation}
\lambda_i = \lambda_i (p^2, \, (p-k)^2,\, k^2 ) \qquad\mbox{ and }\qquad
\tau_i = \tau_i ( p^2, \, (p-k)^2,\, k^2 )  \ ,
\label{TransFF-GapEq}
\end{equation}
with the $\lambda_i$ being the solutions of the WTI. The scalar component of the DSE gives a non-trivial constraint on  some of the transverse form factors 
that is also gauge independent. Note also that  Eq. (\ref{DSE-Fermion-scalar}) is independent of the constant $C$ that appears in the solution of $A(p^2)$.
Eq. (\ref{DSE-Fermion-vector}) depends on $\xi$, suggesting that the constant $C$ is gauge dependent. 
Inserting the functional form for chiral fermions for $A$ into this latter equation, then it can be written as
\begin{equation}
C^2 \, p^2 - C + g^2 \, \mathcal{F}(p^2; \, \xi) = 0
\end{equation}
where $\mathcal{F}(p^2; \, \xi)$ is, in general, a complex number. The solution for $C$ should be independent of $p^2$, which adds again constraints on the form
factors and on $\xi$. It is not clear in what conditions such an equation can be solved and fulfil the requirements just mentioned. Another interpretation of this last
equation is to set $C = 1$ and then it becomes as another integral constraint on the transverse form factors.

\end{appendix}


\end{document}